\documentclass[useAMS]{mn2e}
 \usepackage{times}
 \usepackage{graphicx}
 \usepackage{epsfig}
 \usepackage{amssymb}

 \title[Signalling the trans-Plutonian planets]
       {Extreme trans-Neptunian objects and the Kozai mechanism: signalling the presence of trans-Plutonian planets}

 \author[C. de la Fuente Marcos and R. de la Fuente Marcos]
        {C.~de~la~Fuente~Marcos\thanks{E-mail: nbplanet@fis.ucm.es}
         and
         R. de la Fuente Marcos \\
         Universidad Complutense de Madrid,
         Ciudad Universitaria, E-28040 Madrid, Spain}
 \date{Accepted 2014 June 3.
       Received 2014 June 3;
       in original form 2014 April 23}
 \pubyear{2014}
 
\begin{document}
  \maketitle

  \begin{abstract}
     The existence of an outer planet beyond Pluto has been a matter 
     of debate for decades and the recent discovery of 2012~VP$_{113}$ 
     has just revived the interest for this controversial topic. This 
     Sedna-like object has the most distant perihelion of any known 
     minor planet and the value of its argument of perihelion is close 
     to 0\degr. This property appears to be shared by almost all known 
     asteroids with semimajor axis greater than 150 au and perihelion 
     greater than 30 au (the extreme trans-Neptunian objects or ETNOs), 
     and this fact has been interpreted as evidence for the existence 
     of a super-Earth at 250 au. In this scenario, a population of 
     stable asteroids may be shepherded by a distant, undiscovered 
     planet larger than the Earth that keeps the value of their 
     argument of perihelion librating around 0\degr as a result of the 
     Kozai mechanism. Here, we study the visibility of these ETNOs and 
     confirm that the observed excess of objects reaching perihelion 
     near the ascending node cannot be explained in terms of any
     observational biases. This excess must be a true feature of this 
     population and its possible origin is explored in the framework 
     of the Kozai effect. The analysis of several possible scenarios 
     strongly suggest that at least two trans-Plutonian planets must 
     exist.
  \end{abstract}


  \section{Introduction}
     Are there any undiscovered planets left in the Solar system? The answer to this question is no and perhaps yes! If we are talking about 
     planets as large as Jupiter or Saturn moving in nearly circular orbits with semimajor axes smaller than a few dozen thousand 
     astronomical units, the answer is almost certainly negative (Luhman 2014). However, smaller planets orbiting the Sun well beyond 
     Neptune may exist and still avoid detection by current all-sky surveys (see e.g. Sheppard et al. 2011). Nevertheless, the answer to the 
     question is far from settled and the existence of an outer planet located beyond Pluto has received renewed attention in recent years 
     (see e.g. Gomes, Matese \& Lissauer 2006; Lykawka \& Mukai 2008; Fern\'andez 2011; Iorio 2011, 2012; Matese \& Whitmire 2011). So far, 
     the hunt for a massive trans-Plutonian planet has been fruitless. 

     The recent discovery of 2012 VP$_{113}$ (Sheppard \& Trujillo 2014), a probable dwarf planet that orbits the Sun far beyond Pluto, has 
     further revived the interest in this controversial subject. Trujillo \& Sheppard (2014) have suggested that nearly 250 au from the Sun 
     lies an undiscovered massive body, probably a super-Earth with up to 10 times the mass of our planet. This claim is based on 
     circumstantial evidence linked to the discovery of 2012 VP$_{113}$ that has the most distant perihelion of any known object. The value 
     of the argument of perihelion of this Sedna-like object is close to 0\degr. This property appears to be shared by almost all known 
     asteroids with semimajor axis greater than 150 au and perihelion greater than 30 au (the extreme trans-Neptunian objects or ETNOs), and 
     this has been interpreted as evidence for the existence of a hidden massive perturber. In this scenario, a population of asteroids 
     could be shepherded by a distant, undiscovered planet larger than the Earth that keeps the value of their argument of perihelion 
     librating around 0\degr as a result of the Kozai mechanism. The Kozai mechanism (Kozai 1962) protects the orbits of high inclination 
     asteroids from close encounters with Jupiter but it also plays a role on the dynamics of near-Earth asteroids (Michel \& Thomas 1996). 

     Trujillo \& Sheppard (2014) claim that the observed excess of objects reaching perihelion near the ascending node cannot be the result 
     of observational bias. In this Letter, we study the visibility of extreme trans-Neptunian objects and the details of possible 
     observational biases. Our analysis confirms the interpretation presented in Trujillo \& Sheppard (2014) and uncovers a range of 
     additional unexpected patterns in the distribution of the orbital parameters of the ETNOs. The overall visibility is studied in Section 
     2 using Monte Carlo techniques and an obvious intrinsic bias in declination is identified. The impact of the bias in declination is 
     analysed in Section 3. The distribution in orbital parameter space of real objects is shown in Section 4. In Section 5, our findings 
     are analysed within the context of the Kozai resonance. Results are discussed in Section 6 and conclusions are summarized in Section 7.

  \section{Visibility of trans-Neptunian objects: a Monte Carlo approach}
     Trujillo \& Sheppard (2014) focused their discussion on asteroids with perihelion greater than 30 au and semimajor axis in the range
     150--600 au. Here, we study the visibility of these ETNOs as seen from our planet. The actual distribution of the orbital elements of 
     objects in this population is unknown. In the following, we will assume that the orbits of these objects are uniformly distributed in 
     orbital parameter space. This is the most simple choice and, by comparing with real data, it allows the identification of observational 
     biases and actual trends easily. Using a Monte Carlo approach, we create a synthetic population of ETNOs with semimajor axis, $a\in$ 
     (150, 600) au, eccentricity, $e\in$ (0, 0.99), inclination, $i\in$ (0, 90)\degr, longitude of the ascending node, $\Omega\in$ (0, 
     360)\degr, and argument of perihelion, $\omega\in$ (0, 360)\degr. We restrict the analysis to objects with perigee $<$ 90 au. Out of 
     this synthetic population we single out objects with perihelion, $q = a (1 - e) >$ 30 au, nearly 20 per cent of the sample. Both, the 
     test and Earth's orbit are sampled in phase space until the minimal distance or perigee is found (that is sometimes but not always near 
     perihelion due to the relative geometry of the orbits). Twenty million test orbits were studied. We do not assume any specific size 
     (absolute magnitude) distribution because its nature is also unknown and we are not interested in evaluating any detection efficiency. 
     In principle, our results are valid for both large primordial objects and those (likely smaller) from collisional origin. Further 
     details of our technique can be found in de la Fuente Marcos \& de la Fuente Marcos (2014a, b). 

\begin{figure}
        \centering
         \includegraphics[width=\linewidth]{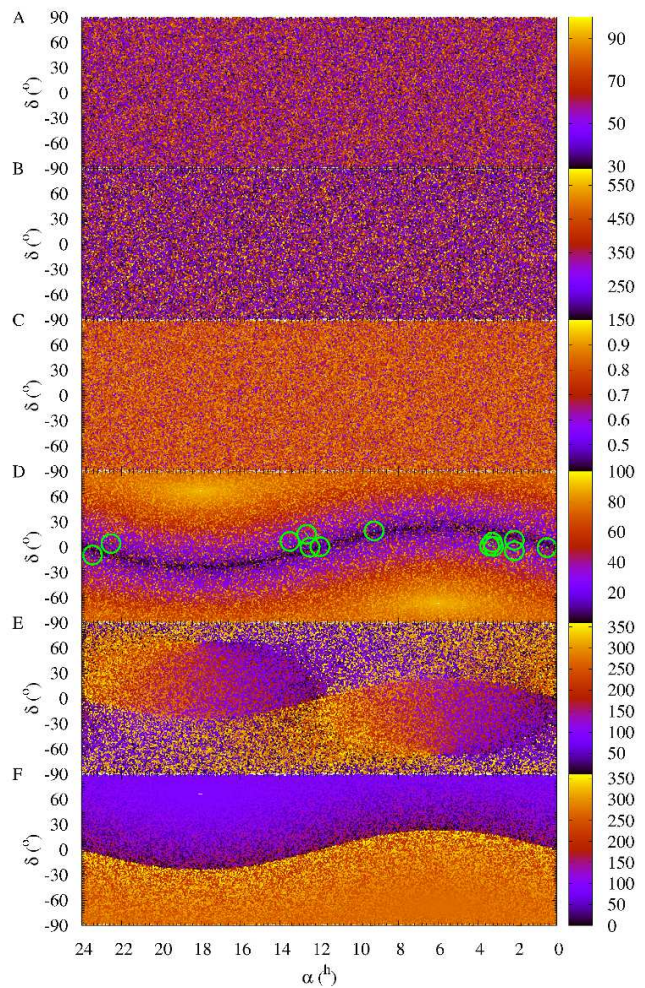}
         \caption{Distribution in equatorial coordinates of the studied test orbits at perigee as a function of various orbital elements and 
                  parameters. As a function of the perigee of the candidate (panel A); as a function of $a$ (panel B); as a function of $e$ 
                  (panel C); as a function of $i$ (panel D); as a function of $\Omega$ (panel E); as a function of $\omega$ (panel F). The
                  associated frequency distributions are plotted in Fig. \ref{full}. The green circles in panel D give the location at 
                  discovery of all the known objects (see Table \ref{discovery}). 
                 }
         \label{hunt2}
      \end{figure}

\begin{figure}
        \centering
         \includegraphics[width=\linewidth]{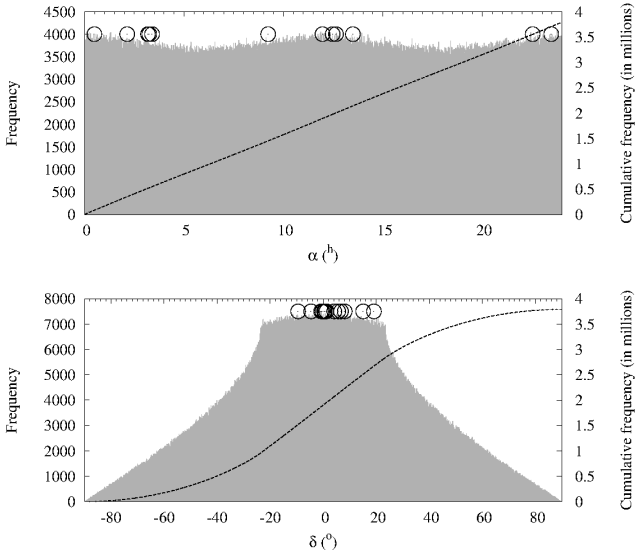}
         \caption{Frequency distribution in equatorial coordinates (right ascension, top panel, and declination, bottom panel) of the 
                  studied test orbits at perigee. The distribution is rather uniform in right ascension and shows a maximum for declinations 
                  in the range -24\degr to 24\degr. The bin sizes are 0.024 h in right ascension and 0\fdg18 in declination, error bars are 
                  too small to be seen. The black circles correspond to objects in Table \ref{discovery}.
                 }
         \label{radec}
      \end{figure}

     Fig. \ref{hunt2} shows the distribution in equatorial coordinates of the studied test orbits at perigee. In this figure, the value of 
     the parameter in the appropriate units is colour coded following the scale printed on the associated colour box. Both perigees (panel 
     A) and semimajor axes (panel B) are uniformly distributed. The eccentricity (panel C) is always in the range 0.4--0.95. Inclination 
     (panel D), longitude of the ascending node (panel E) and argument of perihelion (panel F) exhibit regular patterns. The frequency 
     distribution in equatorial coordinates (see Fig. \ref{radec}) shows a rather uniform behaviour in right ascension, $\alpha$; in 
     contrast, most of the objects reach perigee at declination, $\delta$, in the range -24\degr to 24\degr. Therefore, and assuming a 
     uniform distribution for the orbital parameters of objects in this population, there is an intrinsic bias in declination induced by our 
     observing point on Earth. Under the assumptions made here, this intrinsic bias and any secondary biases induced by it are independent 
     on how far from the ecliptic the observations are made. In other words, if the objects do exist and their orbits are uniformly 
     distributed in orbital parameter space, most objects will be discovered at $|\delta|<$24\degr no matter how complete and extensive the 
     surveys are. But, what is the expected impact of this intrinsic bias on the observed orbital elements?
%
%-------------------------------------------------------------------------------------------------------------------------------------------
%
%
%-------------------------------------------------------------------------------------------------------------------------------------------
%
%
%-------------------------------------------------------------------------------------------------------------------------------------------
%
%
%-------------------------------------------------------------------------------------------------------------------------------------------
%

  \section{The impact of the declination bias}
\begin{figure}
        \centering
         \includegraphics[width=\linewidth]{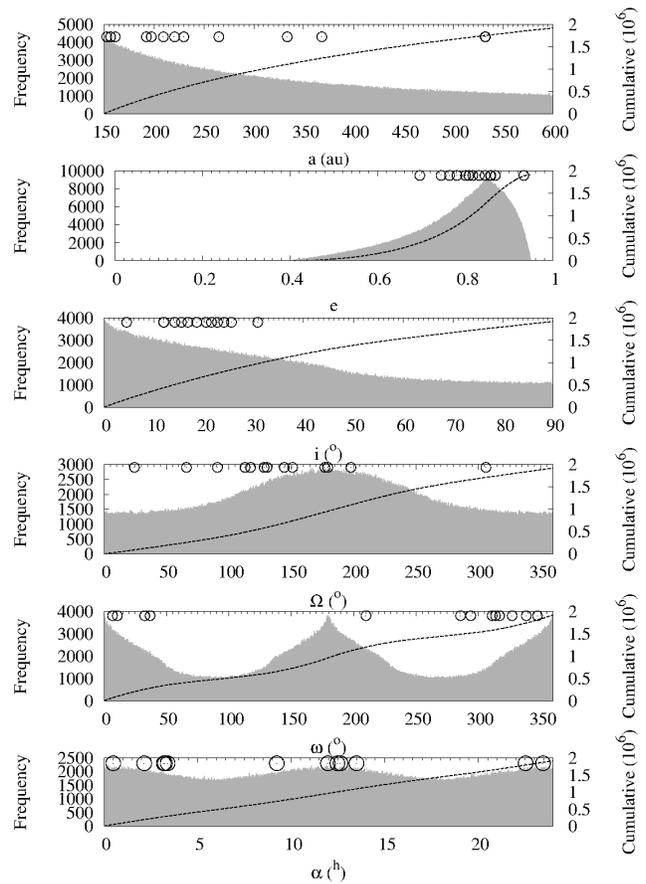}
         \caption{Frequency distribution in right ascension (bottom panel) and the orbital elements of test orbits with $|\delta| <$24\degr. 
                  The bin sizes are 0.45 au in semimajor axis, 0.001 in eccentricity, 0\fdg09 in inclination, 0\fdg36 in longitude of the 
                  node, 0\fdg36 in argument of perihelion and 0.024 h in right ascension, error bars are too small to be seen. The black 
                  circles correspond to objects in Table \ref{discovery}. Data from Table \ref{elements} have been used.
                 }
         \label{rawbias}
      \end{figure}

     So far, all known trans-Neptunian objects with $q>$ 30 au had $|\delta|<$24\degr at discovery which supports our previous result (see 
     Table \ref{discovery} and Fig. \ref{radec}). But this intrinsic bias may induce secondary biases on the observed orbital elements. If 
     we represent the frequency distribution in right ascension and the orbital parameters $a$, $e$, $i$, $\Omega$ and $\omega$ for test 
     orbits with $|\delta| <$24\degr, we get Fig. \ref{rawbias}. Out of an initially almost uniform distribution in $\alpha$, $i$, $\Omega$, 
     and $\omega$ (see Figs \ref{radec} and \ref{full}), we obtain biased distributions in all these four parameters. In contrast, the 
     distributions in $a$ and $e$ are rather unaffected (other than scale factors) by the intrinsic bias in $\delta$. Therefore, most 
     objects in this population should be discovered with semimajor axes near the low end of the distribution, eccentricities in the range 
     0.8--0.9, inclinations under 40\degr, longitude of the ascending node near 180\degr and argument of perihelion preferentially near 
     0\degr and 180\degr. Any deviation from these expected secondary biases induced by the intrinsic bias in declination will signal true 
     characteristic features of this population. Therefore, we further confirm that the clustering in $\omega$ pointed out by Trujillo \& 
     Sheppard (2014) is real, not the result of observational bias. Unfortunately, the number of known objects is small (13), see Table 
     \ref{discovery}, and any conclusions obtained from them will be statistically fragile.
%
%-------------------------------------------------------------------------------------------------------------------------------------------
%
%
%-------------------------------------------------------------------------------------------------------------------------------------------
%

  \section{Distribution in orbital parameter space of real objects}
     The distribution in orbital parameter space of the objects in Table \ref{discovery} shows a number of puzzling features (see Fig. 
     \ref{rawbias}). In addition to the clustering of $\omega$ values around 0\degr (but not 180\degr) already documented by Trujillo \& 
     Sheppard (2014), we observe clustering around 20\degr in inclination and, perhaps, around 120\degr in longitude of the ascending node. 
     The distributions in right ascension, semimajor axis and eccentricity of known objects appear to be compatible with the expectations. 
     However, (90377) Sedna and 2007~TG$_{422}$ are very clear outliers in semimajor axis. Their presence may signal the existence of a very 
     large population of similar objects, the inner Oort cloud (Brown, Trujillo \& Rabinowitz 2004). The distribution in inclination is also 
     particularly revealing. Such a clustering in inclination closely resembles the one observed in the inner edge of the main asteroid belt 
     for the Hungaria family (see e.g. Milani et al. 2010). Consistently, some of these objects could be submitted to an approximate mean 
     motion resonance with an unseen planet. In particular, the orbital elements of 82158 and 2002 GB$_{32}$ are very similar. On the other 
     hand, asteroids 2003 HB$_{57}$, 2005 RH$_{52}$ and 2010 VZ$_{98}$ all have similar $a$, $e$ and $i$, and their mean longitudes, 
     $\lambda$, differ by almost 120\degr (see Table \ref{elements}). This feature reminds us of the Hildas, a dynamical family of asteroids 
     trapped in a 3:2 mean motion resonance with Jupiter (see e.g. Bro\v{z} \& Vokrouhlick\'y 2008). If the three objects are indeed trapped 
     in a 3:2 resonance with an unseen perturber, it must be moving in an orbit with semimajor axis in the range 195--215 au. This 
     automatically puts the other two objects, with semimajor axis close to 200 au, near the 1:1 resonance with the hypothetical planet. 
     Their difference in $\lambda$ is also small (see Table \ref{elements}), typical of Trojans or quasi-satellites. Almost the same can be 
     said about 2003~SS$_{422}$ and 2007~VJ$_{305}$. The difference in $\lambda$ between these two pairs is nearly 180\degr. On the other 
     hand, the clustering of $\omega$ values around 0\degr could be the result of a Kozai resonance (Kozai 1962). An argument of perihelion 
     librating around 0\degr means that these objects reach perihelion at approximately the same time they cross the ecliptic from South to 
     North (librating around 180\degr implies that the perihelion is close to the descending node). When the Kozai resonance occurs at low 
     inclinations, the argument of perihelion librates around 0\degr or 180\degr (see e.g. Milani et al. 1989). At the Kozai resonance, the 
     precession rate of its argument of perihelion is nearly zero. This resonance provides a temporary protection mechanism against close 
     encounters with planets. An object locked in a Kozai resonance is in a metastable state, where it can remain for a relatively long 
     amount of time before a close encounter with a planet drastically changes its orbit.
%
%--------------------------------------------------------------------------------------------------------------------- Discovery coordinates 
%
      \begin{table}
        \centering
        \fontsize{8}{11pt}\selectfont
        \tabcolsep 0.05truecm
        \caption{Equatorial coordinates, apparent magnitudes (with filter if known) at discovery time, absolute magnitude, and $\omega$ for 
                 the 13 objects discussed in this Letter. (J2000.0 ecliptic and equinox. Source: MPC Database.)
                }
        \begin{tabular}{cccccc}
          \hline
             Object             & $\alpha$ ($^{\rm h}$:$^{\rm m}$:$^{\rm s}$) & $\delta$ (\degr:\arcmin:\arcsec) & $m$ (mag)  & $H$ (mag) & $\omega$ (\degr) \\
          \hline
     (82158) 2001 FP$_{185}$    & 11:57:50.69                                 & +00:21:42.7                      & 22.2  (R)  &  6.0      &   6.77           \\
             (90377) Sedna      & 03:15:10.09                                 & +05:38:16.5                      & 20.8  (R)  &  1.5      & 311.19           \\
    (148209) 2000~CR$_{105}$    & 09:14:02.39                                 & +19:05:58.7                      & 22.5  (R)  &  6.3      & 317.09           \\
             2002~GB$_{32}$     & 12:28:25.94                                 & -00:17:28.4                      & 21.9  (R)  &  7.7      &  36.89           \\
             2003~HB$_{57}$     & 13:00:30.58                                 & -06:43:05.4                      & 23.1  (R)  &  7.4      &  10.64           \\
             2003~SS$_{422}$    & 23:27:48.15                                 & -09:28:43.4                      & 22.9  (R)  &  7.1      & 209.98           \\
             2004~VN$_{112}$    & 02:08:41.12                                 & -04:33:02.1                      & 22.7  (R)  &  6.4      & 327.23           \\
             2005~RH$_{52}$     & 22:31:51.90                                 & +04:08:06.1                      & 23.8  (g)  &  7.8      &  32.59           \\
             2007~TG$_{422}$    & 03:11:29.90                                 & -00:40:26.9                      & 22.2       &  6.2      & 285.84           \\
             2007~VJ$_{305}$    & 00:29:31.74                                 & -00:45:45.0                      & 22.4       &  6.6      & 338.53           \\
             2010~GB$_{174}$    & 12:38:29.365                                & +15:02:45.54                     & 25.09 (g)  &  6.5      & 347.53           \\
             2010~VZ$_{98}$     & 02:08:43.575                                & +08:06:50.90                     & 20.3  (R)  &  5.0      & 313.80           \\
             2012~VP$_{113}$    & 03:23:47.159                                & +01:12:01.65                     & 23.1  (r)  &  4.1      & 293.97           \\
          \hline
        \end{tabular}
        \label{discovery}
      \end{table}
%
%-------------------------------------------------------------------------------------------------------------------------------------------
%

  \section{Different Kozai scenarios}
     The most typical Kozai scenario is characterized by the presence of a primary (the Sun in our case), the perturbed body (a massless 
     test particle, an asteroid), and a massive outer or inner perturber such as the ratio of semimajor axes (perturbed versus perturber) 
     tending to zero (for an outer perturber) or infinity (for an inner perturber). In the case of an outer perturber, the critical 
     inclination angle separating the circulation and libration regimes is $\sim$39\degr; for an inner perturber it is $\sim$63\degr (see 
     e.g. Gallardo, Hugo \& Pais 2012). Here, the libration occurs at $\omega$ = 90\degr and 270\degr. Under these circumstances, aphelion 
     (for the outer perturber) or perihelion (for the inner perturber) always occur away from the orbital plane of the perturber. This lack 
     of encounters greatly reduces or completely halts any diffusion in semimajor axis. A classical example of an object submitted to the 
     Kozai effect induced by an outer perturber is the asteroid (3040) Kozai that is perturbed by Jupiter. Another possible Kozai scenario 
     is found when the ratio of semimajor axes (perturbed versus perturber) is close to one. In that case, the libration occurs at $\omega$ 
     = 0\degr and 180\degr; therefore, the nodes are located at perihelion and at aphelion, i.e. away from the massive perturber (see e.g. 
     Milani et al. 1989). Most studies of the Kozai mechanism assume that the perturber follows an almost circular orbit but the effect is 
     also possible for eccentric orbits, creating a very rich dynamics (see e.g. Lithwick \& Naoz 2011). Trujillo \& Sheppard (2014) favour 
     a scenario in which the perturber responsible for the possible Kozai libration experimented by 2012~VP$_{113}$ has a semimajor axis 
     close to 250 au. This puts 2012~VP$_{113}$ near or within the co-orbital region of the hypothetical perturber, i.e. the Kozai scenario 
     in which the ratio of semimajor axes is almost 1. The Kozai mechanism induces oscillations in both eccentricity and inclination 
     (because for them $\sqrt{1-e^{2}} \ \cos{i} = const$) and the objects affected will exhibit clustering in both parameters. This is 
     observed in Fig. \ref{rawbias} but the clustering in $e$ could be the result of observational bias (see above). 

  \section{Discussion}
\begin{figure}
        \centering
         \includegraphics[width=\linewidth]{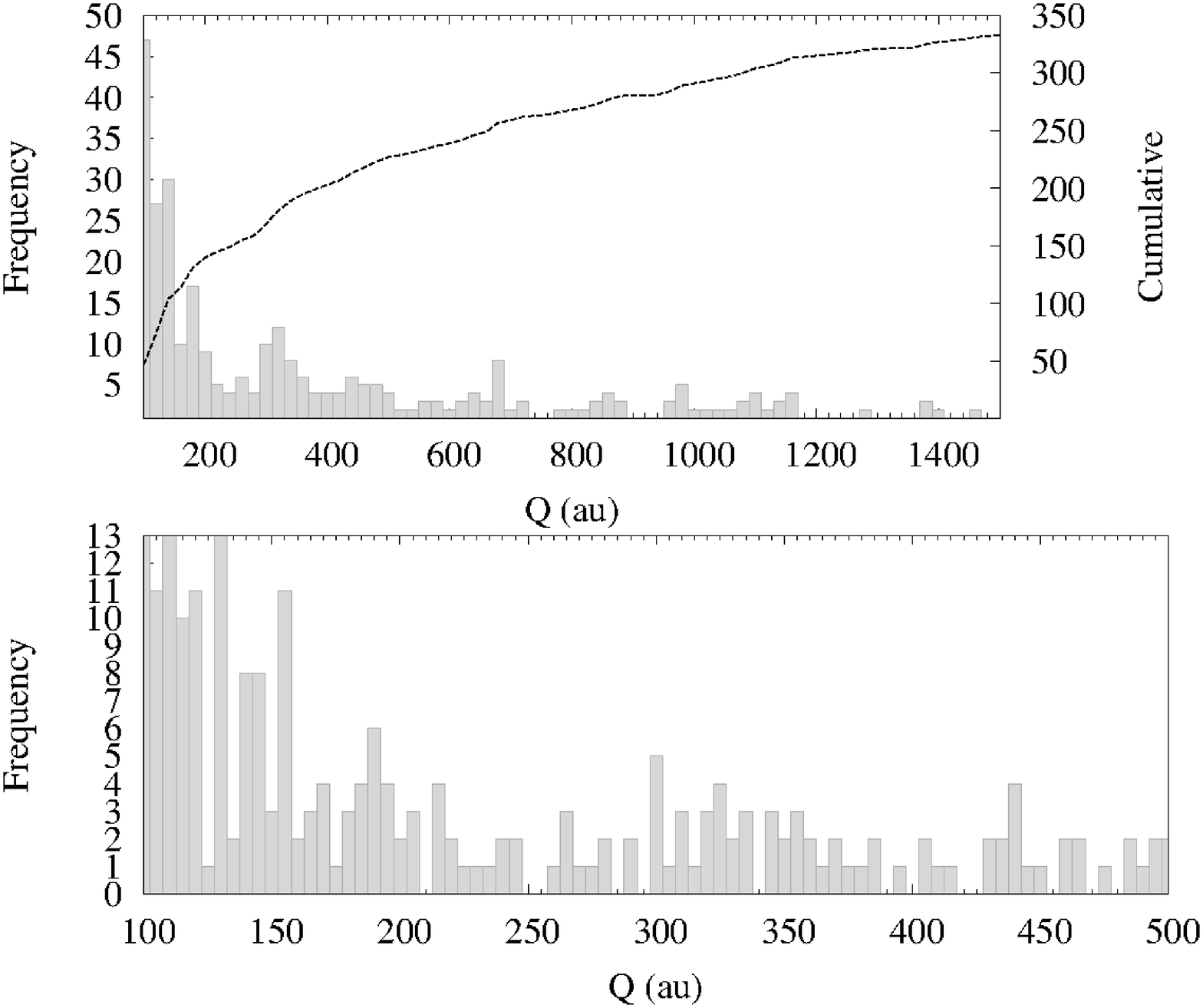}
         \includegraphics[width=\linewidth]{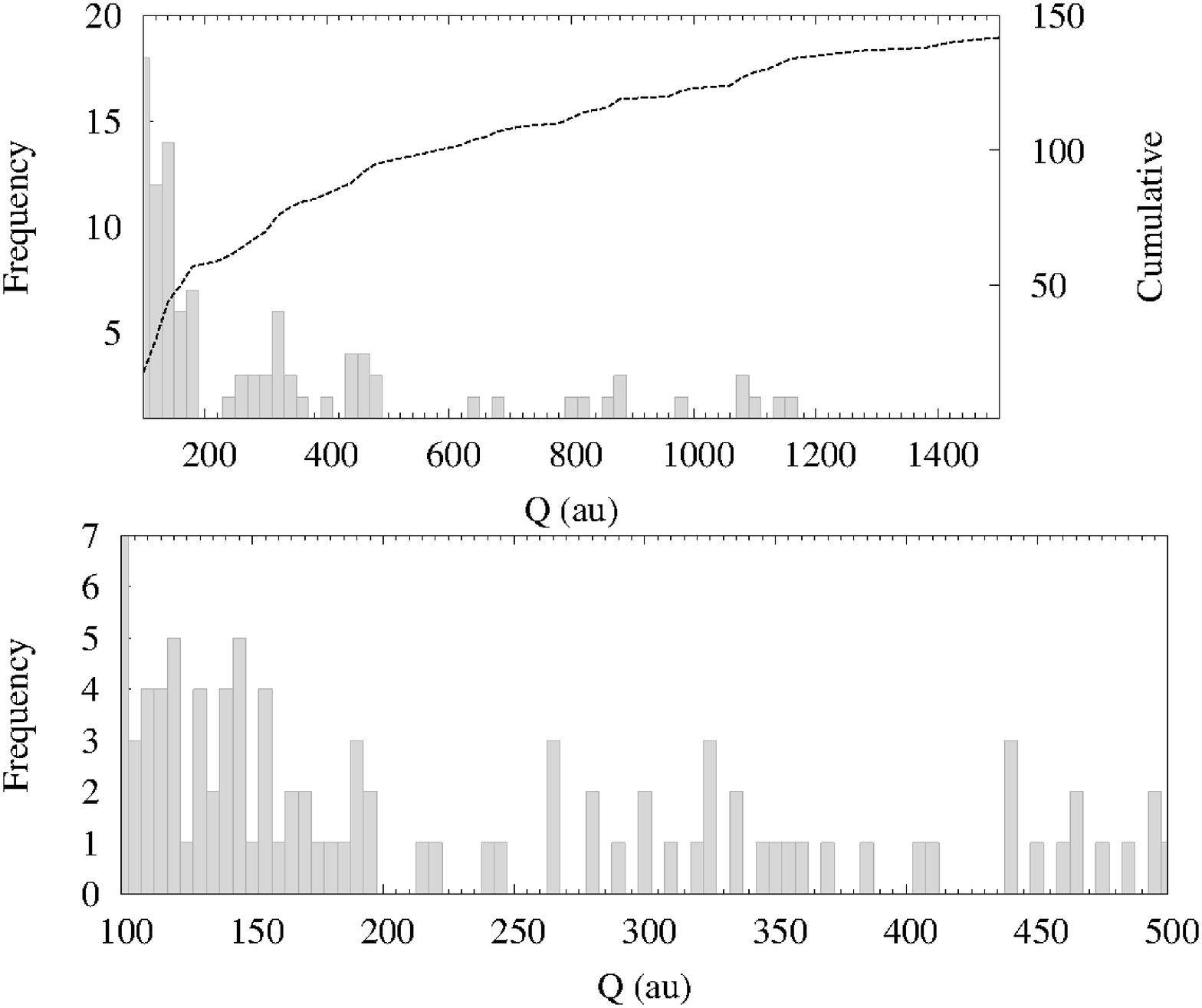}
         \caption{Distribution of aphelia for TNOs and comets with semimajor axis greater than 50 au: all objects (top panels) and only 
                  those with $\omega <$ 35\degr or $\omega >$ 325\degr or $\omega\in$ (145, 215)\degr (bottom panels).  
                 }
         \label{gaps}
      \end{figure}

     Our analysis of the trends observed in Fig. \ref{rawbias} suggests that a massive perturber may be present at nearly 200 au, in 
     addition to the body proposed by Trujillo \& Sheppard (2014). The hypothetical object at nearly 200 au could also be in near resonance 
     (3:2) with the one at nearly 250 au (e.g. if one is at 202 au and the other at 265 au, it is almost exactly 3:2). Any unseen planets 
     present in that region must affect the dynamics of TNOs and comets alike. In this scenario, the aphelia, $Q = a (1 + e)$, of TNOs and 
     comets (moving in eccentric orbits) may serve as tracers of the architecture of the entire trans-Plutonian region. In particular, 
     objects with $\omega\sim$ 0\degr or 180\degr can give us information on the possible presence of massive perturbers in the area because 
     they only experience close encounters near perihelion or aphelion (if the assumed perturbers have their orbital planes close to the 
     Fundamental Plane of the Solar system). Their perihelia are less useful because so far they are $<$ 100 au. However, the presence of 
     gaps in the distribution of aphelia may be a signature of perturbational effects due to unseen planets. Fig. \ref{gaps} shows the 
     distribution of aphelia for TNOs and comets with semimajor axis greater than 50 au. The top two panels show the entire sample. The two 
     panels at the bottom show the distribution for objects with $\omega <$ 35\degr or $\omega >$ 325\degr or $\omega\in$ (145, 215)\degr. 
     These objects have their nodes close to perihelion and aphelion and their distribution in aphelion shows an unusual feature in the 
     range 200--260 au. The number of objects with nodes close to aphelion/perihelion is just four. The total number of objects with 
     aphelion in that range is 18. Immediately outside that range, the number of objects is larger. Although we may think that the 
     difference is significant, it is unreliable statistically speaking because it could be a random fluctuation due to small number 
     statistics. However, a more quantitative approach suggests that the scarcity is indeed statistically significant. If 18 objects have 
     been found in $\omega\in$ (0, 360)\degr, 4 within an interval of 140\degr, and assuming a uniform distribution, we expect 7 objects not 
     4. The difference is just 0.8$\sigma$. Here, we use the approximation given by Gehrels (1986) when $N < 21$: $\sigma \sim 1 + 
     \sqrt{0.75 + N}$. But Fig. \ref{rawbias} indicates that, because of the bias, objects with $\omega$ close to 0\degr or 180\degr are 
     nearly four times more likely to be identified than those with $\omega$ close to 90\degr or 270\degr. So, instead of 7 objects we 
     should have observed 14 but only 4 are found, a difference of 2$\sigma$, that is marginally significant. Therefore, if they are not 
     observed some mechanism must have removed them.
%
%-------------------------------------------------------------------------------------------------------------------------------------------
%
%
%-------------------------------------------------------------------------------------------------------------------------------------------
%
%
%-------------------------------------------------------------------------------------------------------------------------------------------
%
      \begin{figure}
        \centering
         \includegraphics[width=\linewidth]{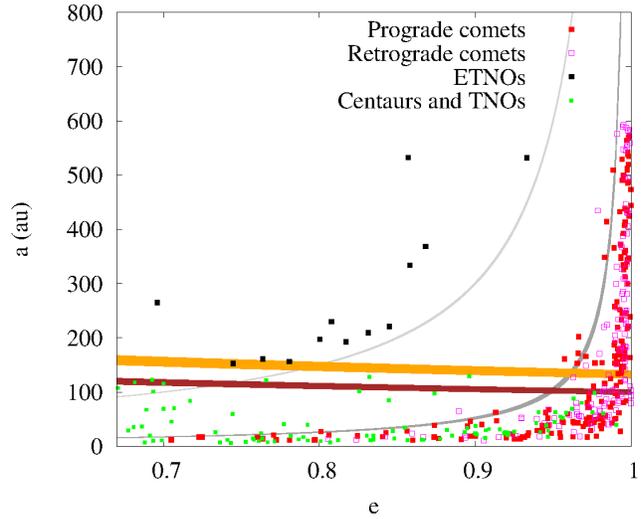}
         \caption{Centaurs, TNOs, ETNOs and comets in the ($e$, $a$) plane. The dark gray area represents the eccentricity/semimajor axis 
                  combination with periapsis between the perihelion and aphelion of Jupiter, the light gray area shows the equivalent 
                  parameter domain if Neptune is considered instead of Jupiter. The brown area corresponds to the ($e$, $a$) combination 
                  with apoapsis between 190 and 210 au and the orange area shows its counterpart for the range 250--280 au.
                 }
         \label{fae}
      \end{figure}
%
%-------------------------------------------------------------------------------------------------------------------------------------------
%

     Fig. \ref{gaps} (top panels) shows an apparent overall decrease in the number of objects with aphelion in the range 200--300 au. The 
     ($e$, $a$) plane plotted in Fig. \ref{fae} confirms that the architecture of that region is unlikely to be the result of a 
     gravitationally unperturbed environment. If there are two planets, one at nearly 200 au and another one at approximately 250 au, their 
     combined resonances may clear the area of objects in a fashion similar to what is observed between the orbits of Jupiter and Saturn but 
     see Hees et al. (2014) and Iorio (2014). On the other hand, it can be argued that ETNOs could be the result of close and distant 
     encounters between the proto-Sun and other members of its parent star cluster early in the history of the Solar system (see e.g. Ida, 
     Larwood \& Burkert 2000; de la Fuente Marcos \& de la Fuente Marcos 2001). However, these encounters are expected to pump up only the 
     eccentricity not to imprint a permanent signature on the distribution of the argument of perihelion in the form of a clustering of 
     values around $\omega$=0\degr. In addition, they are not expected to induce clustering in inclination. 

  \section{Conclusions}
     In this Letter, we have re-examined the clustering in $\omega$ found by Trujillo \& Sheppard (2014) for ETNOs using a Monte Carlo 
     approach. We confirm that their finding is not a statistical coincidence and it cannot be explained as a result of observational bias. 
     Besides, (90377) Sedna and 2007~TG$_{422}$ are very clear outliers in semimajor axis. We confirm that their presence may signal the
     existence of a very large population of similar objects. A number of additional trends have been identified here for the first time:
     \begin{itemize}
        \item Observing from the Earth, only ETNOs reaching perihelion at $|\delta| <$24\degr are accessible.  
        \item Besides clustering around $\omega$ = 0\degr, additional clustering in inclination around 20\degr is observed.
        \item Asteroids 2003 HB$_{57}$, 2005 RH$_{52}$ and 2010 VZ$_{98}$ all have similar orbits, and their mean longitudes differ by 
              almost 120\degr. They may be trapped in a 3:2 resonance with an unseen perturber with semimajor axis in the range 195--215 au. 
        \item The orbits of 82158 and 2002 GB$_{32}$ are very similar. They could be co-orbital to the putative massive object at 195--215 
              au. 
        \item The study of the distribution in aphelia of TNOs and comets shows a relative deficiency of objects with $\omega$ close to 
              0\degr or 180\degr among those with aphelia in the range 200-260 au. The difference is only marginally significant 
              (2$\sigma$), though. Gaps are observed at $\sim$205 au and $\sim$260 au.
     \end{itemize}
     We must stress that our results are based on small number statistics. However, the same trends are found for asteroids and comets, 
     and the apparent gaps in the distribution of aphelia are very unlikely to be the result of Neptune's perturbations or observational 
     bias. Perturbations from trans-Plutonian objects of moderate planetary size may be detectable by the {\it New Horizons} spacecraft 
     (Iorio 2013).

  \section*{Acknowledgements}
     We thank the anonymous referee for her/his helpful and quick report. This work was partially supported by the Spanish `Comunidad de 
     Madrid' under grant CAM S2009/ESP-1496. We thank M. J. Fern\'andez-Figueroa, M. Rego Fern\'andez and the Department of Astrophysics of 
     the Universidad Complutense de Madrid (UCM) for providing computing facilities. Most of the calculations and part of the data analysis 
     were completed on the `Servidor Central de C\'alculo' of the UCM and we thank S. Cano Als\'ua for his help during this stage. In 
     preparation of this Letter, we made use of the NASA Astrophysics Data System, the ASTRO-PH e-print server and the MPC data server.

  \newpage
  \appendix
  \section{Additional figures and tables}

%
%-------------------------------------------------------------------------------------------------------------------------------------------
%
      \begin{figure}
        \centering
         \includegraphics[width=\linewidth]{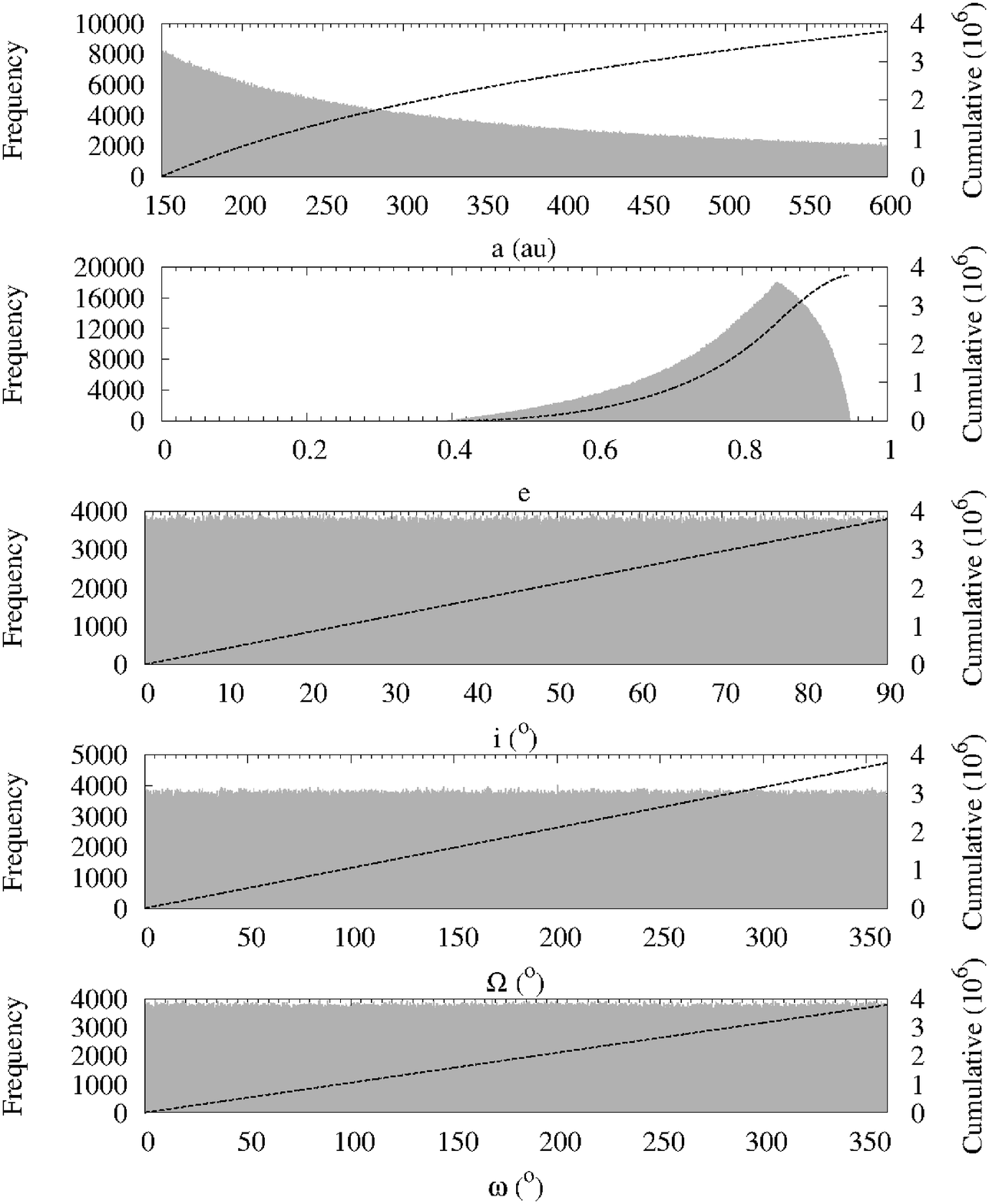}
         \caption{Frequency distributions for the orbital elements of test orbits in Fig. \ref{hunt2}. 
                  Bin sizes are as in Fig. \ref{rawbias}, error bars are too small to be seen. 
                 }
         \label{full}
      \end{figure}
%
%-------------------------------------------------------------------------------------------------------------------------------------------
%
%
%-------------------------------------------------------------------------------------------------------------------------- Orbital elements 
%
      \begin{table*}
        \centering
        \fontsize{8}{11pt}\selectfont
        \tabcolsep 0.35truecm
        \caption{Various orbital parameters ($\varpi = \Omega + \omega$, $\lambda = \varpi + M$) for the 13 objects discussed in this Letter 
                 (Epoch: 2456800.5, 2014-May-23.0 00:00:00.0 UT. J2000.0 ecliptic and equinox. Source: JPL Small-Body Database.)
                }
        \begin{tabular}{cccccccc}
          \hline
             Object             & $a$ (au)    & $e$        & $i$ (\degr) & $\Omega$ (\degr) & $\omega$ (\degr) & $\varpi$ (\degr) & $\lambda$ (\degr) \\
          \hline
     (82158) 2001 FP$_{185}$    & 220.7545067 & 0.84492276 & 30.77926    & 179.32889        &   6.76597        & 186.09486        & 187.24430         \\
             (90377) Sedna      & 532.2664228 & 0.85696250 & 11.92861    & 144.52976        & 311.18801        &  95.71777        &  93.91037         \\
    (148209) 2000~CR$_{105}$    & 229.9196589 & 0.80773939 & 22.70769    & 128.23495        & 317.09262        &  85.32757        &  90.37358         \\
             2002~GB$_{32}$     & 209.4649254 & 0.83128842 & 14.18242    & 177.01044        &  36.88563        & 213.89607        & 213.92324         \\
             2003~HB$_{57}$     & 161.1315216 & 0.76362930 & 15.49540    & 197.85952        &  10.63985        & 208.49937        & 209.44502         \\
             2003~SS$_{422}$    & 197.4196450 & 0.80023290 & 16.80405    & 151.10109        & 209.98241        &   1.08350        &   1.72635         \\
             2004~VN$_{112}$    & 333.5527773 & 0.85809672 & 25.52708    &  66.04930        & 327.23428        &  33.28358        &  33.55408         \\
             2005~RH$_{52}$     & 152.6816879 & 0.74449569 & 20.46892    & 306.19829        &  32.59337        & 338.79166        & 340.86704         \\
             2007~TG$_{422}$    & 531.9002265 & 0.93310126 & 18.57950    & 112.98155        & 285.83713        &  38.81868        &  39.06830         \\
             2007~VJ$_{305}$    & 192.3878720 & 0.81702908 & 11.98914    &  24.38420        & 338.53140        &   2.91560        &   4.04033         \\  
             2010~GB$_{174}$    & 368.2345380 & 0.86809908 & 21.53344    & 130.59114        & 347.52989        & 118.12103        & 121.29941         \\
             2010~VZ$_{98}$     & 156.4583186 & 0.78062638 &  4.50909    & 117.47040        & 313.79473        &  71.26513        &  68.78231         \\
             2012~VP$_{113}$    & 264.9446814 & 0.69599853 & 24.01737    &  90.88555        & 293.97160        &  24.85715        &  27.78384         \\
          \hline
        \end{tabular}
        \label{elements}
      \end{table*}
%
%-------------------------------------------------------------------------------------------------------------------------------------------
%
\end{document}